# Hexagonal Close-packed Skyrmion Lattice in Ultrathin Ferroelectric PbTiO$_3$ Films

Shuai Yuan and Zuhuang Chen*

*School of Materials Science and Engineering, Harbin Institute of Technology, Shenzhen 518055, China and Flexible Printed Electronics Technology Center, Harbin Institute of Technology, Shenzhen 518055, China*

Sergei Prokhorenko, Yousra Nahas and Laurent Bellaiche

*Physics Department and Institute for Nanoscience and Engineering, University of Arkansas, Fayetteville, AR 72701, USA*

Chenhan Liu

*Micro- and Nano-scale Thermal Measurement and Thermal Management Laboratory, School of Energy and Mechanical Engineering, Nanjing Normal University, Nanjing, 210046, P. R. China*

Bin Xu

*Institute of Theoretical and Applied Physics and School of Physical Science and Technology, Soochow University, Suzhou, Jiangsu 215006, China*

Lang Chen

*Department of Physics, Southern University of Science and Technology, Shenzhen, 518055, China*

Sujit Das

*Materials Research Centre, Indian Institute of Science, Bangalore, 560012, India*

Lane W. Martin

*Department of Materials Science and Engineering, University of California, Berkeley, California 94720, USA and Materials Sciences Division, Lawrence Berkeley National Laboratory, Berkeley, California 94720, USA*

*Email: zuhuang@hit.edu.cn

**Abstract:**

Polar skyrmions are topologically stable, swirling polarization textures with particle-like characteristics, which hold promise for next-generation, nanoscale logic and memory. While understanding of how to create ordered polar skyrmion lattice structures and how such structure respond to applied electric fields, temperature, and film thickness remains elusive. Here, using phase-field simulations, the evolution of polar topology and the emergence of a phase transition to a hexagonal close-packed skyrmion lattice is explored through the construction of a temperature-electric field phase diagram for ultrathin ferroelectric $PbTiO_3$ films. The hexagonal-lattice skyrmion crystal can be stabilized under application of an external, out-of-plane electric field which carefully adjusts the delicate interplay of elastic, electrostatic, and gradient energies. In addition, the lattice constants of the polar skyrmion crystals are found to increase with film thickness, consistent with expectation from Kittel's law. Our studies pave the way for the development of novel ordered condensed matter phases assembled from topological polar textures and related emergent properties in nanoscale ferroelectrics.

The topological phases, as defined by its topological invariants, is the new state of matter that has only recently been recognized. Within this field, the topological domain structures in ferroic materials have been known as a source of exotic phenomena [1-10]. Among them, magnetic skyrmions have been extensively studied in recent years for potential applications in low-power and high-efficiency spintronic devices including memories, logic gates, *etc*.[11-15]. Skyrmion crystals (*SkX*), or periodic, lattice-like organizations of skyrmions, in analogy to the vortex phase of type-II superconductors, have been theoretically predicted and experimentally confirmed in several different classes of magnetic materials [16-28]. In contrast to their magnetic counterparts, however, polar skyrmions in ferroelectrics are much less explored [29-32]. Thanks to recent advances in theoretical calculations, synthesis of high-quality thin films, and characterization techniques, polar skyrmions have been observed very recently in [PbTiO$_3$ (PTO)]$_n$/[SrTiO$_3$ (STO)]$_n$ superlattices [33], where exotic emergent properties such as negative permittivity and chirality have been observed [34,35]. Understanding of the evolution of these exotic dipolar textures with temperature *T* and electric field *E* (*i.e.*, the *T-E* phase diagram) remains elusive but is important if researchers want to manipulate these objects for practical use. Even more fundamental and still not demonstrated is if one can actually realize a *SkX* based on these polar textures in ferroelectrics [30]. It is conceivable that a close-packed polar *SkX* could be achieved by further controlling the balance between the charge and lattice degrees of freedom (*i.e.*, electrostatic and elastic boundary conditions); something that would promote additional detail study of these emergent features and their possible collective behaviors [36].

Thus, there is strong motivation to study the *T-E* phase diagram and to explore the formation conditions and phase transition mechanisms for polar *SkX* in ferroelectric materials.

Given the fact that most *SkX* in ferromagnets are stabilized by an external magnetic field, this begs the question as to whether a similar phenomenon could also occur in ferroelectric films under electric fields. Here, phase-field simulations are used to produce a *T-E* phase diagram of polar structures and to explore the possible existence of polar *SkX* in thin films of the prototypical ferroelectric PTO. The simulations reveal that a hexagonal polar *SkX* phase can be induced in a narrow region below the Curie temperature of PTO on the phase diagram. The simulations further demonstrate the dependence of the lattice constant and stability of the *SkX* as a function of film thickness.

Similar to Ref. [33], the PTO films are assumed to grow along the [001] pseudo-cubic direction. An external electric field denoted as $\boldsymbol{E}^{ext} = (0, 0, E_z)$ was applied along [001] (where the *x*, *y*, and *z* axis of Cartesian coordinates correspond to the [100], [010], and [001] pseudo-cubic direction, respectively). According to Landau-Ginzburg-Devonshire theory, the free energy is expressed as $F = \int_V (f_{Land} + f_{grad} + f_{elas} + f_{elec})dV + \int_S f_{surf} dS$, where $f_{Land}$, $f_{grad}$, $f_{elas}$, $f_{elec}$, and $f_{surf}$ represent the bulk Landau potential, gradient, elastic, electrostatic, and surface energy densities, respectively. The electrostatic energy density can be expressed as $f_{elec} = -\boldsymbol{P}_i \boldsymbol{E}_i - \frac{1}{2}\varepsilon_b E_i E_i = -\boldsymbol{P}_i[E_i^{ext} + (1-\theta)\boldsymbol{E}_i^{dep}] - \frac{1}{2}\varepsilon_b[\boldsymbol{E}_i^{ext} + (1-\theta)\boldsymbol{E}_i^{dep}][E_i^{ext} + (1-\theta)\boldsymbol{E}_i^{dep}]$, where $\theta$, $\boldsymbol{E}_i^{ext}$, $\boldsymbol{E}_i^{dep}$, $\varepsilon_b$ are the screening factor, external electric field, depolarization field under open-circuit boundary conditions, and the background permittivity, respectively. As the default conditions, the screening factor $\theta$ is set to 0.6 to achieve an appropriate depolarization field (wherein $\theta = 0$ corresponds to ideal open-circuit conditions and $\theta = 1$ corresponds to ideal short-circuit conditions), and the substrate strain $\varepsilon$ is set to −1.0% to ensure that the polarization has a preferred orientation along the out-of-plane direction. Details of the formalism are provided in the Supplemental Material [37].

An epitaxial (001)-oriented PTO film of size $48\ nm \times 48\ nm \times 6\ nm$ is considered as the initial specimen. As $E_z$ increases from zero at room temperature, the domain structure in the film undergoes a complex topological-phase transition (Fig. 1a). To summarize the evolution of the system as the field is increased from zero to 1.8 MV/cm (top to bottom), we show the real-space distribution of local polarization (Fig. 1a), Pontryagin density (Fig. 1b), and free-energy density (Fig. 1c) of the top surface of the film [37]. The Pontryagin density, as a typical O(3) topological invariant, is used to identify the topological nature of the swirling domain structures [68]. At $E_z = 0$, the PTO films exhibit a labyrinth domain pattern (denoted as the $L$ phase) with meandering features as described before [30,69] (top row, Fig. 1a), where the white arrows and colors denote the direction of in-plane polarization, the saturation of the color represents the polarization magnitude, and the dark regions are domains with out-of-plane polarizations. Similar morphology has been experimentally observed in STO/PTO/STO tri-layer films and [PTO]$_n$/[STO]$_n$ superlattices at room temperature under zero applied field [33,70], indicating the validity of this simulation approach. The Pontryagin density is weak at $E_z = 0$, except slightly larger at the junctions of the labyrinth domain (first row, Fig.1b). The meandering feature of the $L$ phase can be identified clearly from the energy-density map (first row, Fig.1c) wherein the energy-density is high at regions where there is strong out-of-plane polarization (dark regions; first row, Fig.1a), reflecting the huge energy cost due to the depolarization field. As $E_z$ increases, individual skyrmions are first generated at the junctions of the labyrinth domains. When $E_z$ reaches 0.6 MV/cm, local skyrmions emerge and coexist with the labyrinth domains (denoted as the $Sk$ & $L$ phase) (second row, Fig. 1a). The Pontryagin density is larger at each skyrmion core (second row, Fig.1b), and the topological charge of the skyrmion is calculated to be 1 through the surface integration of the Pontryagin density [35]. Correspondingly, the free-energy-density distribution shows that the annular transition region which possesses in-plane polarizations is in a low-energy state (second row, Fig. 1c). Upon further increasing $E_z$ to 1.2 MV/cm, a complete $SkX$ with well-ordered hexagonal, close-packed features is observed (third row, Fig. 1a); analogous to the classical $SkX$ in helical magnets [20,21]. The Pontryagin density also exhibits hexagonal order, with the brightest (largest value) regions corresponding to the positions of the skyrmion cores (third row, Fig. 1b). The hexagonal arrangement is also seen in the energy-density map (third row, Fig. 1c) where the density is large at each skyrmion core, periphery, and the interfacial region between two neighboring skyrmions (where out-of-plane polarization dominates), while it is small at the annular transition region from the core to the periphery of each skyrmion (where in-plane polarization dominates). Finally, when $E_z$ reaches 1.8 MV/cm, the $SkX$ phase is destroyed and transformed into the topologically trivial single-domain phase with pure out-of-plane polarization (denoted as a ferroelectric *FE* phase; fourth row, Fig. 1a), where the Pontryagin density (fourth row, Fig. 1b) is essentially zero and the energy density (fourth row, Fig. 1c) is homogeneously large. Similar skyrmion nucleation and *SkX*-formation were achieved as we varied the temperature under a constant external electric field ($E_z$ = 0.4 MV/cm; Fig. 1d~f). There, as the temperature is increased, the *L* phase observed at 100 K transforms to a mixed *Sk* & *L* phase at 450 K, before fully transforming to a hexagonal, close-packed *SkX* at 500 K, and, ultimately, the domain contrast disappears at ~600 K.

To gain further insights into the nature of the *SkX*, the *SkX* phase was simulated in larger cells, including a PTO film of size $160\ nm \times 160\ nm \times 6\ nm$ at $T = 500$ K. Distinct from the *L* phase at 300 K, the initial state at $E_z = 0$ in the film at higher temperature of 500 K is the stripe domain (denoted as the *S* phase; Supplemental Material [37]). For this same simulation cell, however, increasing the field to $E_z = 0.6$ MV/cm results in a hexagonal *SkX* phase with a periodic array of swirling polar textures over the entire surface (Fig. 2a). The polarization distribution of a single skyrmion in this *SkX* phase is further examined (left column, Fig. 2a) and it is seen that the top and bottom planes of each skyrmion are center-convergent and center-divergent, respectively, while the middle plane has a purely out-of-plane polarization component. That is, the polarization vectors rotate within the plane parallel to the radial direction, indicating that the *SkX* phase is of Néel-type. Noting that the Néel-type *SkX* phase is different from the skyrmion bubble found in PTO/STO superlattices reported in previous studies [33]. Although both are composed of two standard Néel-type skyrmion hats of the top and bottom layers, the connected cylindrical domain wall at the middle layer exhibits different characteristics, wherein the former is an Ising-type wall with out-of-plane polarizations while the latter is Bloch-type with in-plane ones [33]. Such a difference could be due to the larger compressive strain in the current model, which would favor out-of-plane polarization. Besides, the two-point correlation function and the spectra density function have been employed to characterize the ordering of topology in the films quantitatively [37].

To better understand the formation mechanism of the *SkX* phase, a schematic is drawn to analyze the difference and correlation between the domain structures before and after the application of an external electric field (Fig. 2b). Before $E_z$ is applied, for both the *L* and *S* phases, the cross-section view of the domain structures exhibits periodic, long-range-ordered clockwise (CW) and counterclockwise (CCW) polar vortices (Fig. 2bi) [69]. Upon application of a positive $E_z$, the upward polarization region (crimson) is enhanced, and the downward polarization region (indigo) is depressed which results in the adjacent CW-CCW vortices being forced to approach and pair with each other. Ultimately, this results in the formation of Néel-type skyrmions where the polarization at the skyrmion core (periphery) is oriented antiparallel (parallel) to the applied field $E_z$ (Fig. 2bii). The interfacial region between two neighboring skyrmions would, effectively, acts as a topological protection zone. This is supported by the energy-density map in the *SkX* phase shown previously (third row, Fig.1c) since the energy density in the interfacial region is much larger than that near the skyrmion, thus, it would be difficult for the skyrmion to cross the barrier of the interfacial region and destroy the close-packed lattice phase. The close-packed *SkX* lattice phase, in turn, forms from the delicate interplay of elastic, electrostatic, and gradient energies. First, the depolarization field, due to the incomplete screening of the polarization at the surfaces, favors 180° domains (*i.e.*, *L* or *S* domains), which is the basis for the formation of the *SkX* phase. Second, in contrast to the effect of the depolarization field, both the compressive strain and external electric field favor the formation of (uniform) out-of-plane polarization. On the other hand, the gradient energy is mainly related to the polarization inhomogeneity. These energy terms are all in competition and eventually result in the formation of complex polar textures. In particular, the strong competition between the considerable depolarization field caused by incomplete screening and the applied electric field appears to be the most critical to induce the topological-phase transition from stripe domain to the *SkX* phase.

To more completely describe the evolution of the skyrmions under electric field and temperature, we further constructed a phase diagram of the polar textures. Figure 3 shows the *T-E* phase diagram under $E_z$ for a $48\,nm \times 48\,nm \times 6nm$ film. Six characteristic polarization structures are found across the range of temperatures and electric field probed herein, including: *SkX* (Fig. 3a), *L* (Fig. 3b), *S* (Fig. 3c), *Sk glass* (denoting the disordered, sparse skyrmion phase; Fig. 3d), mixed *Sk & L* (Fig. 3e), and mixed *Sk & FE* (Fig. 3f). The *SkX* phase (outlined in white, Fig. 3g) has the absolute value of topological charge of 1 and the *Sk glass* phase corresponds to a topological charge of slightly less than 1. The *S* and *FE* phases have a topological charge of zero, while the *L* phase should (ideally) have zero topological charge, but it can fluctuate between 0 to 0.1 depending on the degree of meandering in the maze. For the mixed-phase regions (*i.e.*, *Sk & FE* and *Sk & L* phases) the topological charge is between 0 and 1. It is interesting to note that an inverse-phase transition takes place as one progresses from the low-temperature *L* phase to the high-temperature *S* phase with increasing temperature. A similar inverse-phase transition has been reported recently in PbZr$_{0.4}$Ti$_{0.6}$O$_3$ ultrathin films using first-principles-based effective Hamiltonian simulations [71]. From the global *T-E* diagram, the *SkX* phase occurs only between 300 K and 500 K. Furthermore, as mentioned above, two different topological-phase transitions exist to reach the *SkX* phase: 1) the *S* to *SkX* transition with increasing electric field (active from 470 to 500 K) and 2) the *L* to *SkX* transition with increasing electric field (active from 300 to 470 K) [37]. Finally, the skyrmions completely disappear for $E_z$ > 2.6 MV/cm or *T* > 550 K, which suggests a transition into the collinear polar state (*i.e.*, the *FE* phase). For the comparative verification of the effectiveness of these results, we conducted simulations *via* a first-principles effective Hamiltonian method to further support the existence of polar *SkX* in PbTiO$_3$ ultrathin films [37]. Besides, we have investigated the general characteristics of *SkX* in a broader system, *e.g.*, tetragonal PbZr$_{1-x}$Ti$_x$O$_3$ (*x*=0.6) films under appropriate boundary conditions [37].

After clarifying the formation mechanism of the hexagonal, close-packed *SkX* phase, the influence of film thickness on the hexagonal lattice constant and stability of the *SkX* phase was further explored. PTO films with size $80\ nm \times 80\ nm \times h\ nm$ at 500 K were selected to simulate the thickness evolution of the polar texture. The simulations reveal that the film thickness $h$ plays a significant role in the polar textures (Fig. 4a). For films with $h < 5$ nm, topologically trivial polar structures, such as in-plane-polarized domains, single domain, or even paraelectric states dominate mainly due to the large depolarization field in ultrathin films [72]. For relatively thicker films, the *SkX* phases emerge and is stabilized within the range 5.0 nm $< h <$ 13.0 nm due to a subtle competition and balance between the depolarization field and external electric field. For a given film thickness, it is found that the distance between two adjacent skyrmion cores (*i.e.*, the modulated period of the *S* phase in Fig. 2bi), remains unchanged with varying external electric fields. Therefore, the lattice constant *w* of the hexagonal polar *SkX* is defined as the distance between the cores of two adjacent skyrmions (Fig. 4b) [73]. Two typical cases of the polarization distributions of *SkX* phases with $h = 11.2$ nm (Fig. 4b) and 6 nm (Fig. 4c) are shown, whereby *w* is found to be 20.5 nm and 13.6 nm, respectively. Generally, topologically trivial domains in ferroelectric, ferromagnetic, and ferroelastic films obey Kittel's law scaling wherein the domain width scales with the square root of film thickness [74-76]. It is interesting to note that the square of the hexagonal lattice constant *w* is found to be linearly proportional to the film thickness *h*, that is, the lattice size of the topological domain structure follows Kittel's scaling law. In thicker films (13 nm $< h <$ 18 nm), the formation of the *SkX* phase becomes challenging due to the weak depolarization field and, instead, the disordered *Sk glass* state becomes dominant in this thickness range.

In summary, phase-field simulations establish a global *T-E* phase diagram of the polar structures and reveal the existence of hexagonal, close-packed Néel-type polar

*SkX* in ferroelectric PTO thin films. The results further demonstrate that the lattice constant of the *SkX* is linearly proportional to the square root of the film thickness, following the Kittel's scaling law. The *SkX* phase in the ferroelectric PTO films is constructed in a periodic lattice manner similar to atomic or molecular crystals, resembling those composed by various topological solitons in other physical contexts, such as the skyrmion lattice in magnetic materials and liquid crystals [77-79], vortex rings in superfluids and ultra-cold gases [80,81], *etc*. These findings open a way for further exploration of ordered condensed-matter phases assembled from polar skyrmions and other topological soliton "building blocks" in nanoscale ferroelectric materials.

## Acknowledgement

This work was funded by the National Natural Science Foundation of China (Grant Nos. U1932116 and 51802057), Guangdong Basic and Applied Basic Research Foundation (Grant No. 2020B1515020029), Shenzhen Science and Technology Innovation project (Grant No. JCYJ20200109112829287), and Shenzhen Science and Technology Program (Grant No. KQTD20200820113045083). Z.H.C. has been supported by "the Fundamental Research Funds for the Central Universities" (Grant No. HIT.OCEF.2022038) and "Talent Recruitment Project of Guangdong" (Grant No. 2019QN01C202). S.Y. acknowledges the financial support from China Postdoctoral Science Foundation (No. DB24407028), and Guangxi Science and Technology Program (Grant No. AD20159059). S.P., Y.N. and L.B. are grateful for the support by the Vannevar Bush Faculty Fellowship (VBFF) Grant No. N00014-20-1-2834 from the Department of Defense, and the Army Research Office under the ETHOS MURI Grant W911NF-21-2-0162. The first-principles based effective Hamilton simulations were performed the University of Arkansas with the support from the Arkansas High Performance Computing Center (AHPCC). B. Xu acknowledges financial support from National Natural Science Foundation of China under Grant No. 12074277. C. L. acknowledges the financial support from the Natural Science Foundation of Jiangsu Province (Grant no. BK20210565). S.D. acknowledges Science and Engineering Research Board (SRG/2022/000058) and Indian Institute of Science start-up grant for financial support. L.W.M. acknowledges support from the Army Research Office under the ETHOS MURI via cooperative agreement W911NF-21-2-0162.

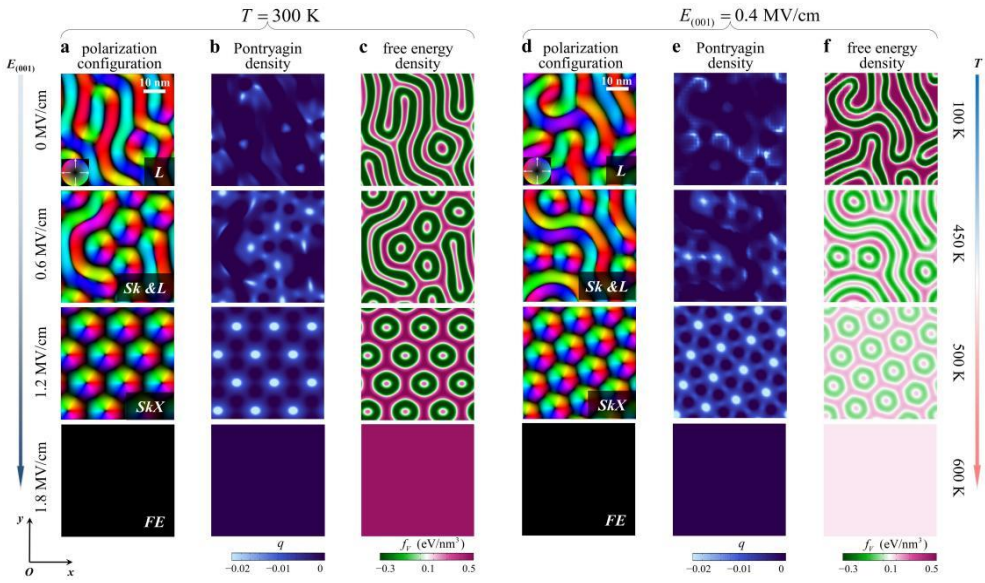

**FIG. 1 (color online). Variation of polar texture with electric field and temperature in a PbTiO₃ film mimicked by size of 48 nm × 48 nm × 6 nm.** Evolution of (a) polarization configuration, (b) Pontryagin density, and (c) free energy density of the top layer of the film with increment of an electric field at $T = 300$ K. d)-f) The topological transition with increment of temperature at a fixed electric field $E_z = 0.4$ MV/cm.

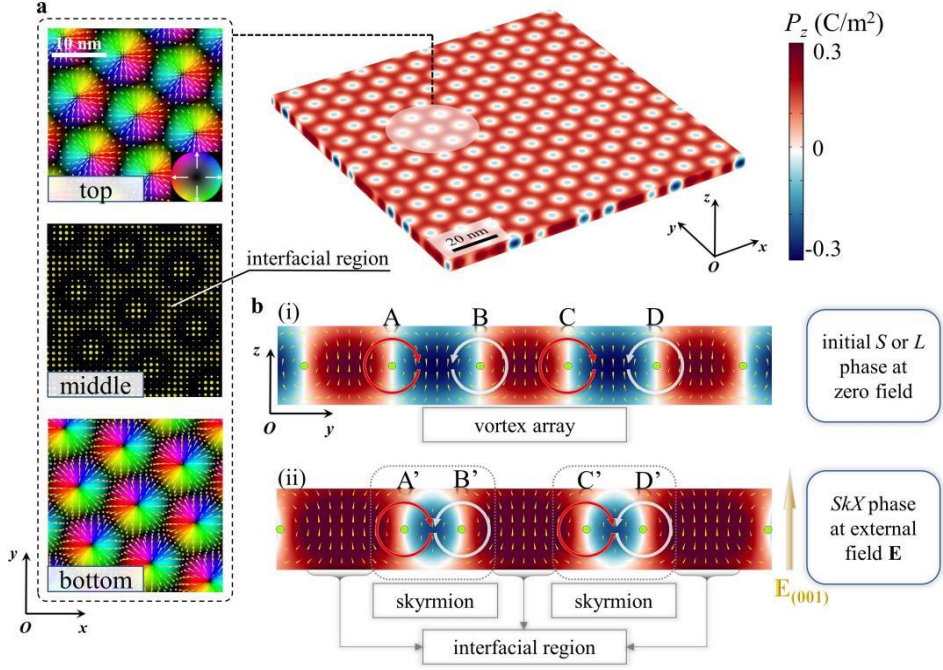

**FIG. 2 (color online). Nature of the *SkX* and its formation mechanism.** (a) The *SkX* in a PbTiO$_3$ film of 160 nm × 160 nm × 6 nm. The plane view of the $P_z$ distribution of *SkX* is on the right, and the polarization distributions on the top layer, middle layer, and bottom layer of the white-marked region are on the left. The golden arrow indicates the polarization vector. (b) Schematic of *SkX* formation mechanism from the cross-sectional perspective of the film. The green dots indicate vortex cores. Crimson and indigo represent region with polarizations upward and downward, respectively.

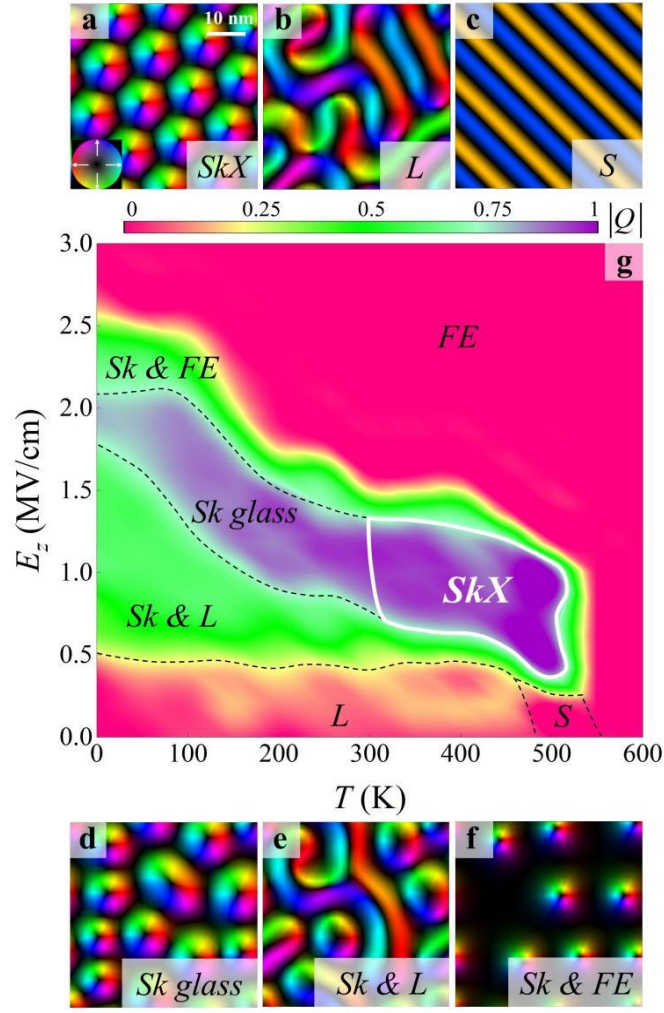

**FIG. 3 (color online). *T-E* phase diagram of polar texture in a 48 *nm* × 48 *nm* × 6*nm* PTO film, the color bar denotes the absolute value of the topological charge $Q$.** Six typical domain structures are shown around the phase diagram (g), including (a) *SkX*, (b) parallel-stripe domain (*S*), (c) labyrinth-stripe domain (*L*), (d) the disordered sparse skyrmions (*Sk glass*), (e) mixed phase of *Sk glass* and labyrinth-stripe domain (*Sk & L*), (f) mixed phase of *Sk glass* and single domain (*Sk & FE*).

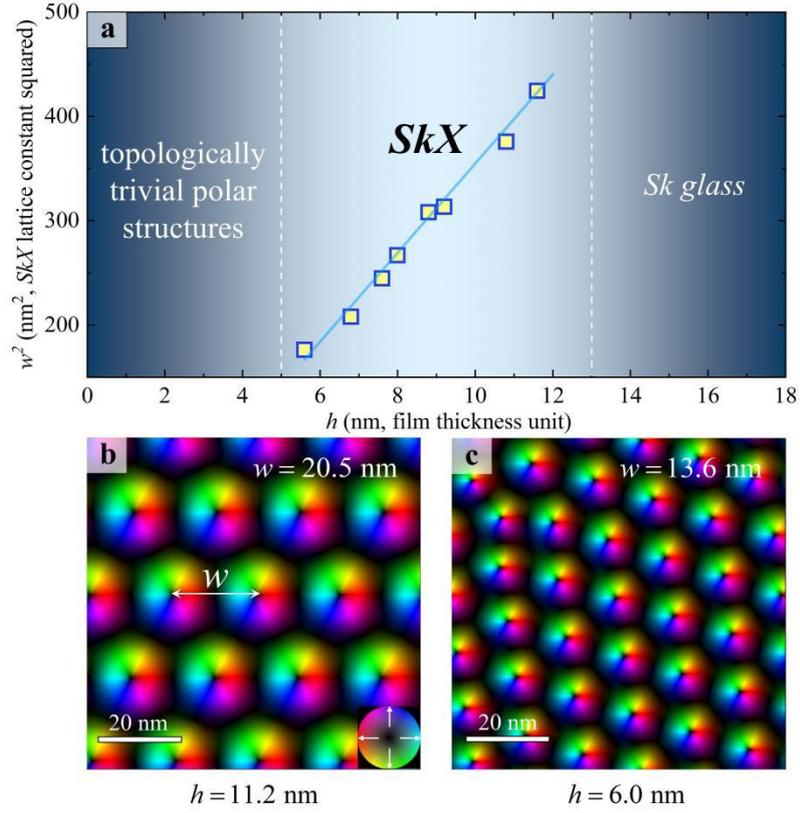

**FIG. 4 (color online). Kittel's law in the *SkX* phase.** (a) Thickness dependence of the polar structures in films of $80\ nm \times 80\ nm \times h$. The straight line is a fit to Kittel's law. Polarization distribution of the *SkX* phases for film thickness $h$ of (b) 11.2 nm and (c) 6.0 nm.